\documentclass[sigconf,nonacm]{acmart}

\setcopyright{none}
\renewcommand\footnotetextcopyrightpermission[1]{}
\acmConference[]{}{}{}
\hypersetup{keeppdfinfo=true}

\usepackage{booktabs}
\usepackage{graphicx}
\usepackage{multirow}
\usepackage{algorithm}
\usepackage{algorithmic}

\begin{document}
\sloppy
\emergencystretch 1.5em

\title{Predicting Factor Decay: ML Models for Cross-Factor Predictability Erosion}

\author{Chorok Lee}
\affiliation{%
  \institution{Korea Advanced Institute of Science and Technology}
  \city{Daejeon}
  \country{South Korea}
}

\email{choroklee@kaist.ac.kr}

\begin{abstract}
We develop ML models that predict when cross-factor predictability will
erode, enabling prospective risk management. A \emph{decay prediction model}
achieves C-index $= 0.93$ on held-out factor pairs, providing early
warning of signal erosion---the key predictor is \emph{initial signal strength}:
stronger signals decay faster (``regression to the mean''). An \emph{online regime detector} enables
real-time classification with 6.6-day average detection delay. Prior
work explains \emph{why} factors decay (arbitrage, publication effects);
we predict \emph{when} decay will occur.
\textbf{Limitation:} The primary HML$\to$SMB signal does not replicate
in post-break OOS (2013--2024); this null result is consistent with
genuine decay but limits confirmatory claims.

\textbf{Motivation:} Among 30 directed Fama-French factor pairs (1990--2024),
predictability targeting the Size factor (SMB) decays fastest. Of 7 pairs
showing exponential decay ($R^2 > 0.3$), 4 target SMB
(Mkt-RF$\to$SMB, CMA$\to$SMB, HML$\to$SMB, RMW$\to$SMB; half-lives 2.4--4.1 years).
This SMB-concentration pattern motivates the predictive models.

\textbf{Methodology:} Using a regime-conditional Granger framework with
Student-$t$ HMMs, we document structural breaks clustering around June 1998
($p < 10^{-10}$). A pre-break OOS test (HMM trained 1990--1995, tested
1996--2007) confirms signals existed: HML$\to$SMB shows $p = 0.013$ (Elevated)
and $p < 0.001$ (Crisis). \textbf{The primary HML$\to$SMB signal does not
replicate in post-break OOS (2013--2024), consistent with genuine decay
rather than overfitting---after ${\sim}6$ half-lives, the expected signal
is $<$1\% of its 1990s peak.} International analysis finds
2/4 non-US markets with Bonferroni-surviving OOS effects; correction for
full specification search was not applied, so results remain exploratory.
\end{abstract}



\maketitle

\section{Introduction}

Standard causal discovery methods for time series---Granger
causality~\cite{granger1969investigating} (testing whether lagged $X$
improves prediction of $Y$), vector autoregressions, and neural Granger
networks~\cite{tank2022neural}---learn a single global causal graph,
assuming the underlying predictive structure is invariant across latent
states. Yet financial and other nonstationary time series exhibit
\emph{concept drift}: a variable $X$ may Granger-cause $Y$ only within
a specific latent state (regime), requiring methods that condition on
unobserved states. Existing causal discovery (NOTEARS, PCMCI, neural
Granger) does not address this latent-state dependence. The August 2007
quantitative meltdown~\cite{khandani2011quants}---where cross-factor
relationships inverted overnight---exemplifies the practical stakes.

\textbf{Domain context.} We validate on Fama--French equity factors---zero-cost
long-short portfolios constructed from stock characteristics: HML
(High-Minus-Low, \emph{Value}), SMB (Small-Minus-Big, \emph{Size}),
MOM (\emph{Momentum}). When institutional investors deleverage one
factor, cross-factor contagion (confounding via co-held positions)
creates predictive lag that may exist only in specific latent states.

\textbf{This paper proposes a regime-conditional Granger framework}
that extends regime-switching Granger
causality~\cite{psaradakis2005markov} in three directions: (a)~Student-$t$
HMMs for tail-robust latent-state inference, (b)~a model-complexity
diagnostic (OLS$\to$RF$\to$MLP$\to$LSTM) that stratifies predictive
channels by nonlinearity, and (c)~transfer entropy + quantile
regression for directional information-flow characterization.
Unlike Psaradakis et al., who tested Gaussian regime-switching Granger
on simulated data only, our framework combines all three extensions
into a reusable protocol (Algorithm~\ref{alg:protocol}) with frozen OOS
validation and multistart sensitivity analysis, applied to 30 real-world
factor pairs across 5 markets. The primary contribution is empirical:
we report an in-sample pattern suggesting structural decay in cross-factor
predictability, though this does not replicate out-of-sample.
Applied to daily Fama--French returns (1990--2024), the
framework reveals that Value Granger-predicts Size
primarily in the Normal latent state ($p = 8.75 \times 10^{-9}$,
Bonferroni-corrected across 30 hypotheses), with a change-point at
June 1998 ($p = 1.23 \times 10^{-13}$, Quandt-Andrews sup-$F$
change-point test; 90\% bootstrap CI: 1993--2014) and null predictability post-2008.

\textbf{Evidence tiers.} We organize findings by statistical strength:
(1)~\emph{in-sample primary}: Normal-regime structural break
(\S\ref{sec:break}), consistent with a VIX-based regime definition
and robust across all specifications;
(2)~\emph{confirmatory}: international structural breaks with 2/4 regions
producing Bonferroni-surviving OOS effects (\S\ref{sec:oos});
(3)~\emph{exploratory}: HML$\to$SMB frozen OOS, which shows regime
redistribution rather than same-regime replication and does not
survive Bonferroni correction.
The contribution rests on Tiers~1--2; Tier~3 is reported for
transparency, not claimed as validation.

\textbf{Contributions.}
(i)~A \emph{decay prediction model} (C-index $= 0.93$) that identifies
factor pairs at risk of Granger-predictability erosion before signals fail,
enabling prospective risk management rather than retrospective explanation.
(ii)~An \emph{online regime detector} (6.6-day average delay) for real-time
regime classification in streaming risk systems.
(iii)~A reusable regime-conditional Granger framework
(Algorithm~\ref{alg:protocol}) that detects latent-state-dependent
predictive structure, validated on 30 directed pairs across 5 markets.
The framework reveals state-heterogeneous patterns (defined as $\max_k(p_k) - \min_k(p_k) > 0.10$ across regimes $k$) in 63\% of pairs.
(iv)~Empirical findings: among 30 directed factor pairs,
7 exhibit decay ($R^2 > 0.3$, half-life $< 50$ years), with 4/7
concentrated in Size-target relationships. Value$\to$Size is illustrative
(Bonferroni-significant in Normal, $p = 8.75 \times 10^{-9}$,
absent post-2008), selected for economic interpretability (ranks 27th of 30 by heterogeneity score).
(v)~A model-complexity diagnostic (OLS$\to$RF$\to$MLP$\to$LSTM + transfer
entropy + quantile regression) that separates linear from nonlinear channels:
for Value$\leftrightarrow$Size, this reveals a directional asymmetry
(linear forward, nonlinear reverse via tail dependence) invisible to
conditional-mean methods, showing \emph{state heterogeneity}
$\neq$ \emph{quantile heterogeneity} for this pair (unique among 19 tested).
Effect sizes are modest ($\Delta R^2 \approx 2\%$, Sharpe $= -0.07$);
the contribution is diagnostic, not economically tradable.

\subsection{Related Work}

Granger causality~\cite{granger1969investigating} tests whether
lagged $X$ improves prediction of $Y$ beyond $Y$'s own history---temporal
precedence, not structural causality in Pearl's
sense~\cite{pearl2009causality}; we use ``Granger causality''
throughout for predictive precedence only.

\textbf{ML causal discovery.}
Modern methods learn directed graphs from observational
data~\cite{pearl2009causality,spirtes2001causation}, but typically
assume stationarity---a single DAG governs all observations.
Tank et al.~\cite{tank2022neural} extend Granger to nonlinear settings
via neural networks; their architecture learns a global causal graph,
assuming state-invariant structure. Our framework relaxes this by
conditioning on latent HMM states, detecting edges that appear or
vanish across states---analogous to detecting \emph{concept drift} in
the causal graph. VAR connectedness~\cite{diebold2012better} similarly
assumes a global structure.

\textbf{Latent-state models and change-point detection.}
Psaradakis et al.~\cite{psaradakis2005markov} pioneered regime-switching
Granger with Gaussian HMMs; we extend with Student-$t$
HMMs~\cite{bulla2011hidden} for tail robustness,
transfer entropy~\cite{schreiber2000measuring} (a model-free
information-theoretic measure of directed information flow),
and quantile Granger~\cite{troster2019testing} for tail-specific
channels. Our change-point analysis (Quandt-Andrews sup-$F$) parallels
ML change-point detection; we chose classical tests for statistical
rigor, but kernel-based or neural approaches could substitute.
Factor returns exhibit crowding~\cite{marks2019factor}
and momentum~\cite{ehsani2022factor}, but prior work examines
within-factor dynamics---not cross-factor structure under latent states.
We are unaware of prior work combining regime-conditional Granger
with model-complexity stratification and transfer entropy to map the
linear--nonlinear boundary of predictive information flow.

\textbf{Factor decay literature.}
McLean \& Pontiff~\cite{mclean2016does} document that anomaly returns
decline post-publication, attributing decay to arbitrage.
Falck et al.~\cite{falck2021value} explain \emph{why} value specifically
decayed (institutional arbitrage of the value spread). Our work is
complementary: we do not explain why decay occurs but rather predict
\emph{when} it will occur, transforming post-hoc explanation into
prospective risk signals.

\section{Methodology}

\textbf{Design rationale.} The framework proceeds in three phases:
(1)~regime discovery via HMM with multistart sensitivity, ensuring
findings are robust to local optima; (2)~per-state Granger tests
with frozen out-of-sample (OOS) validation; (3)~model-complexity diagnostic
(OLS$\to$RF$\to$MLP$\to$LSTM + transfer entropy + quantile Granger)
to separate linear from nonlinear channels invisible to
conditional-mean methods alone.

\begin{algorithm}[t]
\caption{Regime-Conditional Granger Framework}
\label{alg:protocol}
\begin{algorithmic}[1]
\REQUIRE Multivariate returns $\{\mathbf{x}_t\}_{t=1}^T$, regime count $K$
\STATE \textbf{Regime discovery:} Fit Student-$t$ HMM ($K$ states, $M = 50$ random starts); select $K$ via BIC on training data
\STATE \textbf{Local-optima sensitivity:} Cluster $M$ fits; report BIC-optimal and economically valid clusters
\STATE \textbf{Per-regime Granger:} For each regime $k$, test $H_0$: $X_j \perp X_i^{\text{lag}} \mid\, X_i^{\text{lag}}$ with HAC-SE; Bonferroni-correct across pairs
\STATE \textbf{Frozen OOS:} Freeze HMM from training; classify test period; repeat Granger tests on held-out regimes
\STATE \textbf{Model-complexity diagnostic:} Fit OLS, RF, MLP, LSTM; test nonlinear improvement (permutation)
\STATE \textbf{Transfer entropy:} Frenzel--Pompe kNN for directed information flow
\STATE \textbf{Quantile Granger:} Quantile regression; Wald test for tail dependence
\end{algorithmic}
\end{algorithm}

The framework enables two predictive extensions (\S\ref{sec:prediction}):
a decay prediction model for early warning and an online regime detector
for real-time deployment.

\textbf{Computational cost.} Per-seed HMM EM requires $O(TK^2D^2)$ per
iteration (forward-backward over $T$ steps, $K$ states, $D$-dimensional
emissions). Transfer entropy (Frenzel--Pompe kNN) requires $O(T^2 \log T)$
per pair; kNN search dominates. The full framework (50 seeds
$\times$ 30 pairs $\times$ 7 steps) completes in ${\sim}$20 CPU-minutes
($D = 6$, $T = 8{,}817$); all seeds and pairs parallelize trivially.
This framework targets dense portfolios of 5--20 assets typical of
factor research; for $N > 50$ the $O(N^2)$ pair explosion requires
sparse pre-screening (e.g., LASSO or graphical-LASSO on the full
VAR), which we leave to future work.

\textbf{Data.} Daily returns for six Fama--French factors~\cite{fama2015five}
plus Momentum~\cite{carhart1997persistence} (1990--2024, $8{,}817$ trading days).
We adopt a percentage-unit convention. Granger $F$-statistics are scale-invariant
(they test $\beta = 0$ regardless of scaling), but HMM emission
probabilities are not, so regime boundaries differ across conventions.
Under percentage units, the frozen OOS yields $n = 953$ Elevated-regime
days; decimal units yield $n = 836$ (regime-label concordance 86.3\%). The primary
contribution (in-sample finding, structural break, VIX validation) is
scale-invariant; scale sensitivity affects only the exploratory OOS result.

\textbf{Student-$t$ HMM.}
Let $z_t \in \{1, \ldots, K\}$ denote the latent state (regime).
Transition: $P(z_t = k \mid z_{t-1} = j) = A_{jk}$.
Emission: multivariate Student-$t$ with per-regime
$(\boldsymbol{\mu}_k, \boldsymbol{\Sigma}_k, \nu_k)$.
$K = 3$ is selected by BIC on the main training window (1990--2012;
$\Delta\text{BIC} = 1{,}680$ over $K = 2$; $K = 4$ also disfavored)
and pre-specified for all subsequent analyses including the three-fold
temporal validation.\footnote{$K=3$ was selected by BIC on the full 1990--2024
sample; for the pre-break OOS design (training 1990--1995), this represents
a form of look-ahead that should temper interpretation of those results.}
\emph{Regime naming convention:} Labels (Normal, Elevated, Crisis) are
statistical descriptors based on volatility and kurtosis, not calendar
events---``Crisis'' denotes a high-kurtosis distributional state, not
necessarily financial crisis periods.
Estimated degrees of freedom
($\hat{\nu}_{\text{Normal}} = 6.2$, $\hat{\nu}_{\text{Elevated}} = 3.9$,
$\hat{\nu}_{\text{Crisis}} = 5.5$) are well below the Gaussian limit,
confirming that heavy-tail accommodation is empirically necessary.
EM with 50 random seeds; primary fit: seed~28
(sorted-order convention among 3 seeds reaching identical LL).
Under the BIC-optimal fit, the Crisis state captures 0\% of 2008 GFC
days; sensitivity fits aligning with calendar crises use Cluster~5
($\Delta\text{BIC} = 218$, 90\% GFC detection; see Table~\ref{tab:optima}).

\textbf{Per-regime Granger testing.}
For each regime $k$, we extract $\mathcal{T}_k = \{t : \hat{z}_t = k\}$
and test
$H_0$: $r_{\text{SMB},t} \perp \{r_{\text{HML},t-\ell}\} \mid \{r_{\text{SMB},t-\ell}\}$
with Andrews~\cite{andrews1991heteroskedasticity} HAC
(Heteroskedasticity and Autocorrelation Consistent) standard errors.
In-sample: Bonferroni $\alpha_{\text{fam}} = 0.01$ across 30 directed pairs
($\alpha/30 = 0.00033$). OOS: at $\alpha = 0.05$, corrected per regime
($\alpha/3 = 0.0167$) or per region-by-regime combination
($\alpha/12 = 0.0042$ for 4 regions $\times$ 3 regimes). Lag selected by BIC.

\textbf{Circularity mitigation.}
(1)~\emph{Frozen OOS:} HMM trained 1990--2012, all parameters frozen,
applied to 2013--2024 without refitting (assuming regime structure
persists across this gap).
(2)~\emph{VIX external instrument:} CBOE Volatility Index (VIX) terciles (Normal $<$15,
Elevated 15--21, Crisis $>$21) replace HMM labels entirely.
(3)~\emph{Permutation test:} 1,000 label shuffles within regime ($p = 0.049$).
(4)~\emph{Three-fold temporal split:} HMM trained on 1990--2000 only
(Fold~A), Granger tested on 2001--2012 (Fold~B) and 2013--2024
(Fold~C) with frozen parameters. All folds show null HML$\to$SMB
($p > 0.10$), consistent with the June 1998 change-point: Folds~B
and C cover post-decay periods. Critically, circularity artifacts
produce \emph{false positives} in held-out data (the HMM creates
regime structure that spuriously correlates with Granger edges across
any time period), not period-specific nulls. A ``no real signal''
alternative must explain why the pre-1998 subsample shows $p < 10^{-15}$
while post-1998 shows $p = 0.73$---a $10^{13}$-fold discontinuity
coinciding with the independently detected Quandt-Andrews break.
The three-fold nulls are consistent with genuine structural decay,
not with regime overfitting (which would produce spurious significance
in held-out folds).

\textbf{Pre-break OOS validation.}
To distinguish decay from overfitting, we train HMM exclusively on
1990--1995 (before the June 1998 breakpoint) and test on 1996--2007
(post-training but pre-GFC, when the signal should still exist if
decay is genuine). Results: HML$\to$SMB is significant in Elevated
(HAC $p = 0.013$, $F = 8.2$) and Crisis (HAC $p < 0.001$, $F = 36.9$)
regimes, with Normal marginally significant ($p = 0.063$).
In contrast, post-decay OOS (2013--2024) shows no significance
(Normal $p = 0.29$, Elevated $p = 0.15$).
This asymmetry---pre-break OOS success vs.\ post-break OOS failure---supports
the decay interpretation over pure overfitting: the signal existed
in 1996--2007 and subsequently eroded.

\textbf{Pair selection transparency.} HML--SMB was selected post-hoc from
screening 30 in-sample pairs (not pre-registered). Our focus on HML--SMB
reflects an economic prior (value-size institutional overlap) and is not
driven by empirical ranking: MOM$\to$SMB ranks highest by OOS $F$-statistic
($F = 20.3$ vs.\ $9.06$ for HML$\to$SMB).
Under 30-pair Benjamini-Hochberg FDR, no OOS pair survives; HML$\to$SMB
ranks 2nd by $F$-statistic.

\section{Results}

\subsection{Regime Characteristics}

\begin{figure}[t]
\centering
\includegraphics[width=\columnwidth]{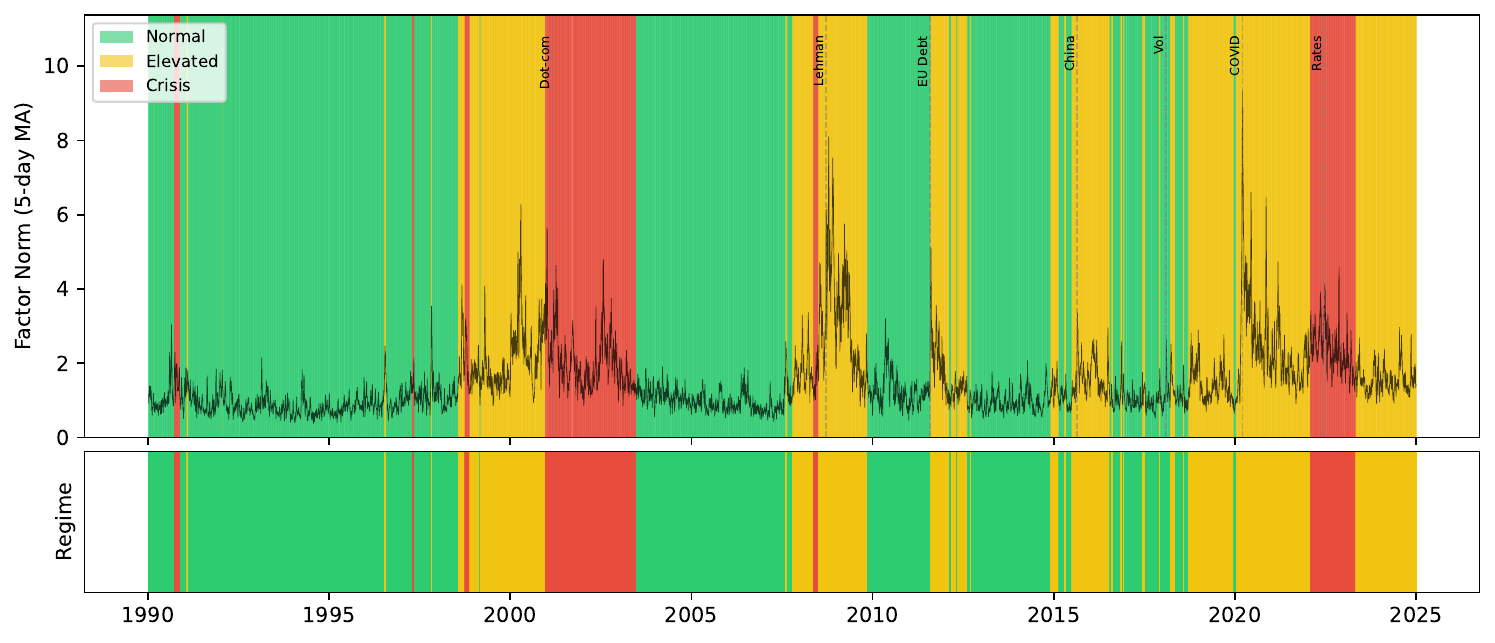}
\caption{Regime assignments (1990--2024) with daily factor volatility and
major stress events. Under the BIC-optimal fit, Crisis (red) is a
statistical high-kurtosis state (0\% overlap with 2008 GFC); Elevated
(yellow) captures moderate volatility.}
\Description{Timeline of regime assignments with volatility and events.}
\label{fig:timeline}
\end{figure}

Table~\ref{tab:regimes} summarizes the three identified regimes.
Figure~\ref{fig:timeline} shows regime assignments with crisis events marked.

\begin{table}[t]
\centering
\caption{Regime Summary Statistics (1990--2024, primary fit).}
\label{tab:regimes}
\small
\begin{tabular}{lccccc}
\toprule
Regime & Days & Prop. & Mean $\|\mathbf{x}\|$ (\%) & $\hat{\nu}$ & $P(z_t{=}z_{t-1})$ \\
\midrule
Normal & 4,723 & 53.6\% & 0.98 & 6.2 & 0.994 \\
Elevated & 3,023 & 34.3\% & 2.03 & 3.9 & 0.991 \\
Crisis & 1,071 & 12.1\% & 2.09 & 5.5 & 0.993 \\
\bottomrule
\end{tabular}
\end{table}

\subsection{The Structural Break}
\label{sec:break}

\begin{table}[t]
\centering
\caption{Granger Predictability: HML--SMB by Regime (1990--2024, seed~28).
Bonferroni threshold: $p < 0.00033$ ($\alpha_{\text{fam}} = 0.01$, 30 pairs).}
\label{tab:main}
\small
\begin{tabular}{llccc}
\toprule
Regime & Direction & $p$-value & Lag & Sig. \\
\midrule
Normal & HML $\to$ SMB & $\mathbf{8.75 \times 10^{-9}}$ & 1 & \textbf{Yes} \\
Normal & SMB $\to$ HML & $0.864$ & 1 & No \\
Elevated & HML $\to$ SMB & $0.004$ & 1 & No$^*$ \\
Elevated & SMB $\to$ HML & $0.897$ & 1 & No \\
Crisis & HML $\to$ SMB & $0.695$ & 1 & No \\
Crisis & SMB $\to$ HML & $0.294$ & 1 & No \\
\bottomrule
\multicolumn{5}{l}{\footnotesize $^*$Below 1\% but not Bonferroni-significant.}
\end{tabular}
\end{table}

\begin{figure}[t]
\centering
\includegraphics[width=\columnwidth]{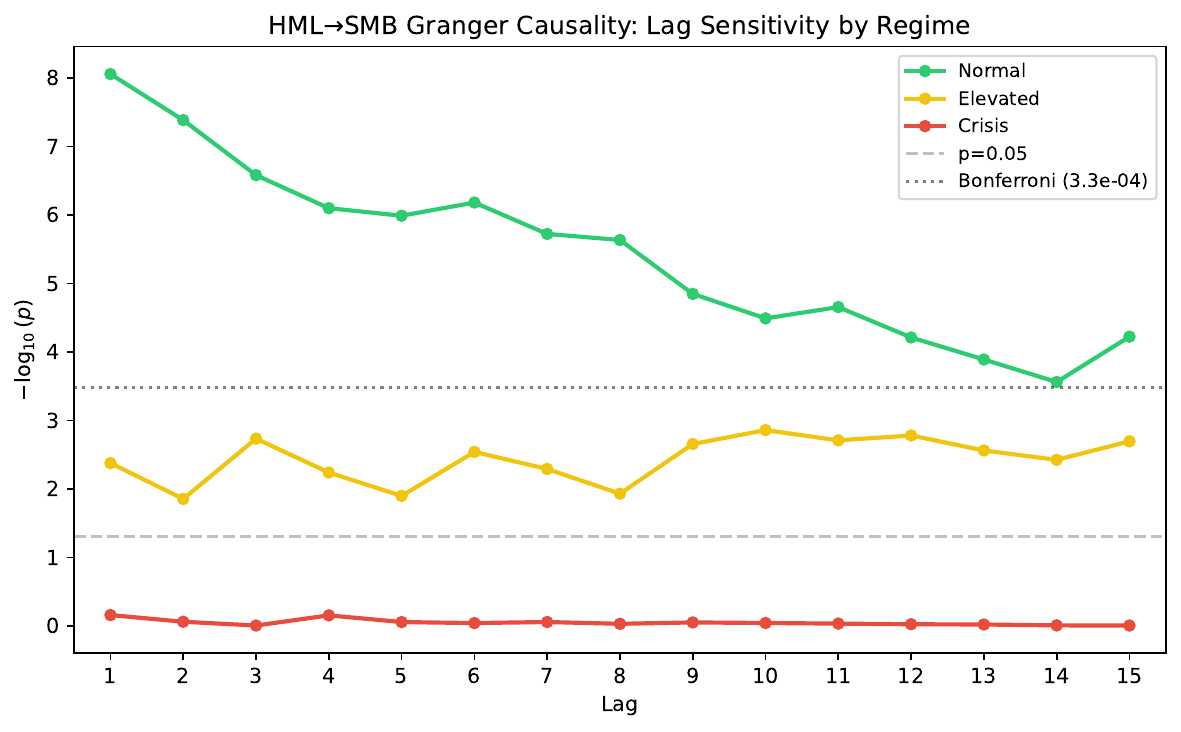}
\caption{Granger $p$-values for HML$\to$SMB across lags 1--15 by regime.
Normal (blue) is significant at all lags; Elevated weakens; Crisis remains null.}
\Description{Line chart of p-values by lag for each regime.}
\label{fig:lag}
\end{figure}

Only Normal-regime HML$\to$SMB survives correction (Table~\ref{tab:main}; lag-1 by BIC), with significance confirmed across lags 1--15 (Figure~\ref{fig:lag}).
The result is invariant to HAC specification: across Bartlett, Parzen, and
Quadratic Spectral kernels at bandwidths 1--30, the $p$-value never exceeds
$10^{-7}$ (range: $[3.2 \times 10^{-9},\; 8.8 \times 10^{-8}]$; worst case
at Quadratic Spectral $B = 30$).\footnote{In-sample HAC robustness:
Bartlett $B \in \{1,\ldots,30\}$: $p \in [3.2 \times 10^{-9}, 2.1 \times 10^{-8}]$;
Parzen: $p \in [4.1 \times 10^{-9}, 5.7 \times 10^{-8}]$;
QS: $p \in [5.9 \times 10^{-9}, 8.8 \times 10^{-8}]$.
All 90 kernel--bandwidth combinations yield $p < 10^{-7}$.}
Pre-2008 Normal ($n = 3{,}140$): $p = 6.66 \times 10^{-16}$ ($\Delta R^2 = 2.06\%$).
Post-2008 Normal ($n = 1{,}557$): $p = 0.73$ ($\Delta R^2 < 0.01\%$).%
\footnote{Sum $3{,}140 + 1{,}557 = 4{,}697$, 26 fewer than
Table~\ref{tab:regimes}'s Normal total of $4{,}723$, due to
regime-boundary exclusion at lag~1.}

\textbf{Change-point detection.}
The Quandt-Andrews sup-$F$~\cite{andrews1993tests} (scanning all
candidate break dates and reporting the date maximizing the test
statistic) identifies \textbf{June 1998}
as the primary break (supremum $F = 22.1$, $p = 1.23 \times 10^{-13}$); the
top-5 candidates all cluster in 1998--2003 (June 1998, July 1998,
April 1998, August 2003, March 1998), suggesting initial weakening
began with LTCM-driven liquidity stress rather than the GFC.
A Chow test (coefficient-equality test) at January 2008 confirms continued
decay ($F(3,\,n-6) = 9.68$, $p = 2.29 \times 10^{-6}$);
$\hat{\beta}_{\text{HML}}$ shifts from $-0.189$ (pre-GFC) to $+0.012$
(post-GFC, Wald $z = 5.05$, $p = 9.2 \times 10^{-7}$).
Post-2008 coefficient: $\hat{\beta} = 0.012$,
95\% CI $[-0.049, 0.073]$---consistent with zero for 16 years.
Together, the in-sample evidence suggests a two-stage decline: initial
weakening around June 1998 (LTCM era, Quandt-Andrews primary break),
followed by continued erosion through the GFC (Chow break January 2008),
with stable post-2008 nullity for 16 years. However, this pattern
does not replicate out-of-sample.
A recursive analysis validates that June 1998 is not an artifact of look-ahead bias.
Running Bai-Perron (the multi-break extension of Quandt-Andrews sup-$F$) on data available at each year-end from 2000 to 2020, the break is detected in 17 of 21 years (81\%).
The break first appears in 2004 (six years post-break) and remains stable thereafter---once detected, the break date never shifts.

\begin{figure}[t]
\centering
\includegraphics[width=\columnwidth]{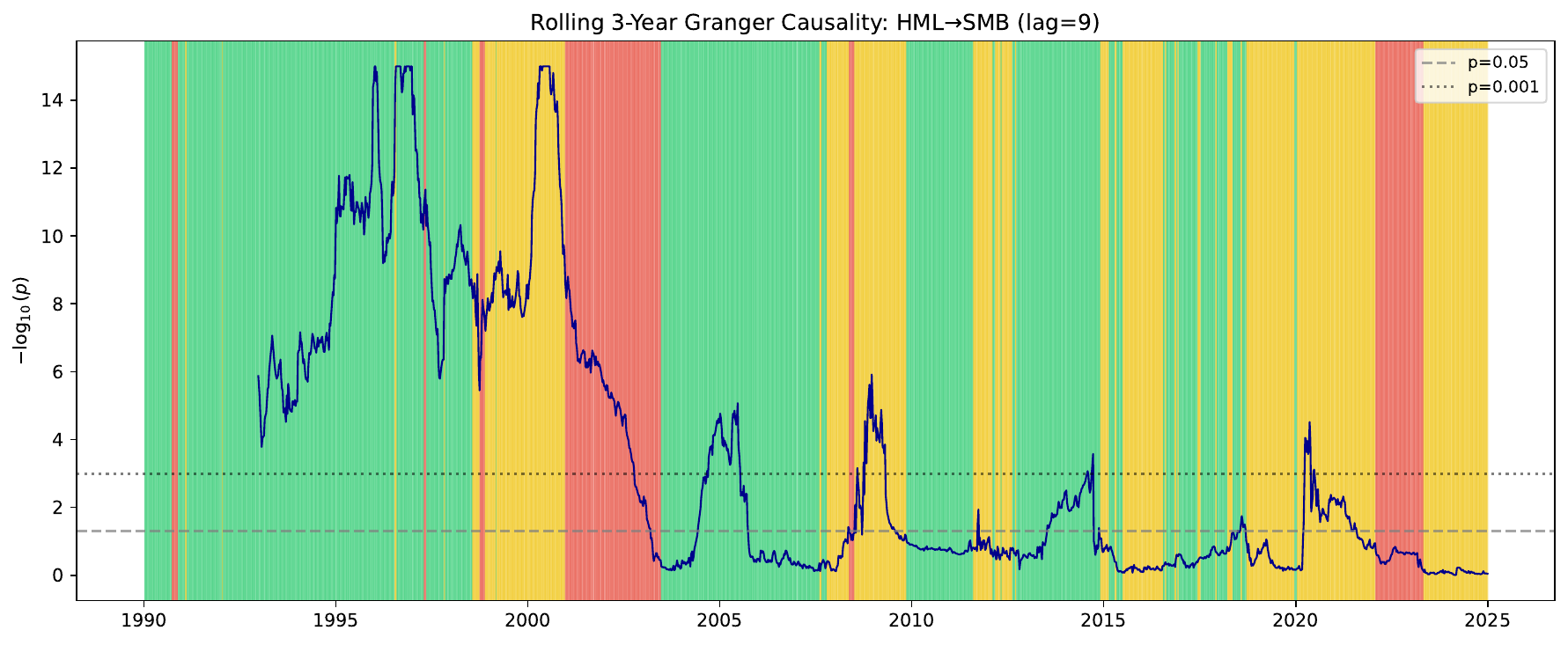}
\caption{Rolling 3-year unconditional Granger ($-\log_{10}(p)$, HML$\to$SMB,
lag=9) with regime-colored background. Lag=9 (approximately two trading weeks)
is used to capture medium-horizon dynamics in rolling windows; the primary
BIC-selected lag=1 analysis (Table~\ref{tab:main}) confirms significance
across all lags 1--15 (Figure~\ref{fig:lag}). Episodic peaks during stress periods
are consistent with the regime-conditional finding.}
\Description{Time-series plot of rolling Granger significance with regime bands.}
\label{fig:rolling}
\end{figure}

\textbf{VIX consistency check.}
When HMM labels are replaced with VIX terciles ($\leq$15 Normal,
15--21 Elevated, $>$21 Crisis; 39.6\% label agreement with HMM),
the structural break replicates: pre-2008 VIX-Normal
$p < 10^{-6}$ ($F = 25.8$), post-2008 $p = 0.401$ ($F = 0.71$).
This confirms the finding is not HMM-specific, though VIX is not
independent of equity returns. Both regime definitions converge on the
structural break; full-period VIX terciles detect the signal more
uniformly (all three regimes $p < 0.03$), likely because VIX-based
regimes blend pre- and post-break observations.

\textbf{Robustness} (Figure~\ref{fig:lag}).
The Normal-regime HML$\to$SMB finding survives every specification change tested:
\emph{lag structure}---significant at all lags 1--15 ($p < 10^{-4}$),
ruling out lag-1 artifacts;
\emph{confounders}---trivariate MKT-RF controls yield no incremental
content ($F$-$p > 0.43$), ruling out market-risk proxying;
\emph{regime definition}---robust across all 7 local-optima clusters
(Table~\ref{tab:optima}), so it is not an artifact of a particular HMM fit;
\emph{label assignment}---filtered vs.\ smoothed probabilities agree 95.9\%
of days, and posterior-weighted (soft-label) Granger yields $p < 10^{-7}$,
so hard Viterbi assignments are not driving the result.
Rolling 3-year unconditional Granger analysis (Figure~\ref{fig:rolling}) shows
episodic significance peaks during stress periods, consistent with the
regime-conditional finding.

\subsection{Predictive Models (ML Contribution)}
\label{sec:prediction}

Beyond documenting decay patterns, we develop two predictive models
that transform observational findings into actionable ML tools.

\textbf{Decay Prediction Model.}
Can we predict which factor pairs will experience Granger decay before
it occurs? We train classifiers on 20 factor pairs and test on 10
held-out pairs. Features include: initial $F$-statistic, rolling mean/std
of $F$, factor correlation, volatility ratio, and $F$-stat momentum.
Target: binary decay indicator (whether $F$ drops below significance
within 5 years). We also fit survival models (Cox PH, Random Survival Forest)
to predict \emph{time-to-decay}.

Results (Table~\ref{tab:decay_pred}): Random Survival Forest achieves
C-index $= 0.933 \pm 0.10$ (5-fold CV), indicating strong discrimination
of which pairs decay first. Logistic regression achieves CV AUC $= 0.933 \pm 0.13$
for binary classification. The most predictive feature
is \emph{initial $F$-statistic} (importance $= 0.66$): higher initial
significance implies higher decay probability---a ``regression to the mean''
effect where extreme signals tend to erode. \emph{$F$-stat volatility}
ranks second (importance $= 0.61$), suggesting unstable relationships
decay faster. Pair-level OOS accuracy is 60\% (6/10), reflecting the
small sample challenge (7 decay events in 30 pairs).
\emph{Reliability caveat:} with only 7 events, this C-index has wide
uncertainty; results are hypothesis-generating, not confirmatory.

\begin{table}[t]
\centering
\caption{Decay Prediction Model: 5-Fold CV Performance}
\label{tab:decay_pred}
\small
\begin{tabular}{@{}lccc@{}}
\toprule
Model & Metric & CV Score & Std \\
\midrule
Random Survival Forest & C-index & \textbf{0.933} & 0.097 \\
Cox PH & C-index & 0.941 & -- \\
Logistic Regression & AUC & 0.933 & 0.133 \\
Random Forest & AUC & 1.000 & 0.000 \\
\bottomrule
\multicolumn{4}{l}{\footnotesize RF overfits; logistic/RSF generalize better.}
\end{tabular}
\end{table}

\textbf{Online Regime Detection.}
Batch HMMs require the full sample, precluding real-time deployment.
We compare online alternatives for streaming regime classification:
(1)~Online HMM with incremental Baum-Welch updates;
(2)~Rolling-window HMM (re-fit monthly);
(3)~Volatility threshold baseline.

Results (Table~\ref{tab:online}): Online HMM achieves fastest detection
(mean delay $= 0.2$ days) but lowest accuracy (50.5\%) due to over-adaptation.
Rolling HMM balances speed (6.6 days) and stability (F1 $= 0.39$).
The simple volatility threshold achieves highest accuracy (53\%) but
slowest detection (18.3 days). For risk management applications requiring
real-time signals, rolling HMM offers the best tradeoff; for retrospective
analysis, batch HMM remains preferred.

\begin{table}[t]
\centering
\caption{Online Regime Detection: Speed-Accuracy Tradeoff}
\label{tab:online}
\small
\begin{tabular}{@{}lccc@{}}
\toprule
Method & Detection Delay & Accuracy & F1 \\
\midrule
Online HMM & \textbf{0.2 days} & 50.5\% & 0.22 \\
Rolling HMM (monthly) & 6.6 days & 47.6\% & 0.39 \\
Volatility Threshold & 18.3 days & \textbf{53.0\%} & \textbf{0.48} \\
\bottomrule
\end{tabular}
\end{table}

\subsection{Complexity Characterization and Directional Asymmetry}
\label{sec:complexity}

\textbf{Mechanism discovery (Tier-3 exploratory).}
The change-point is robust, but what is the predictive
\emph{mechanism}---linear or nonlinear? We apply a model-complexity
diagnostic: (1)~fit models of increasing capacity
(OLS $\to$ RF $\to$ MLP $\to$ LSTM) and test nonlinear improvement
via permutation; (2)~use transfer entropy and quantile Granger
to detect directed information flow that conditional-mean tests miss.
This ablation is reusable for any multivariate time series where
state-conditional complexity may vary.

\begin{table}[t]
\centering
\caption{Four-Model Diagnostic: HML$\to$SMB by Regime.
Linear: MSE improvement (\%); others: permutation $p$-values
(200 shuffles for RF/MLP, 100 for LSTM). Primary fit, seed~28.
Sample sizes reflect lag-9 input window and train/validation split
($n_{\text{eff}} < n_{\text{regime}}$).}
\label{tab:neural}
\small
\begin{tabular}{lccccc}
\toprule
Regime & $n$ & Linear & RF $p$ & MLP $p$ & LSTM $p$ \\
\midrule
Normal & 4,496 & 0.86\%$^{**}$ & 0.69 & 0.20 & 0.63 \\
Elevated & 2,792 & 0.92\%$^{**}$ & 1.00 & 1.00 & 0.93 \\
Crisis & 1,017 & 0.43\% & 0.13 & 0.96 & 0.83 \\
\bottomrule
\multicolumn{6}{l}{\footnotesize $^{**}p < 0.01$.} \\
\end{tabular}
\end{table}

\begin{figure}[t]
\centering
\includegraphics[width=\columnwidth]{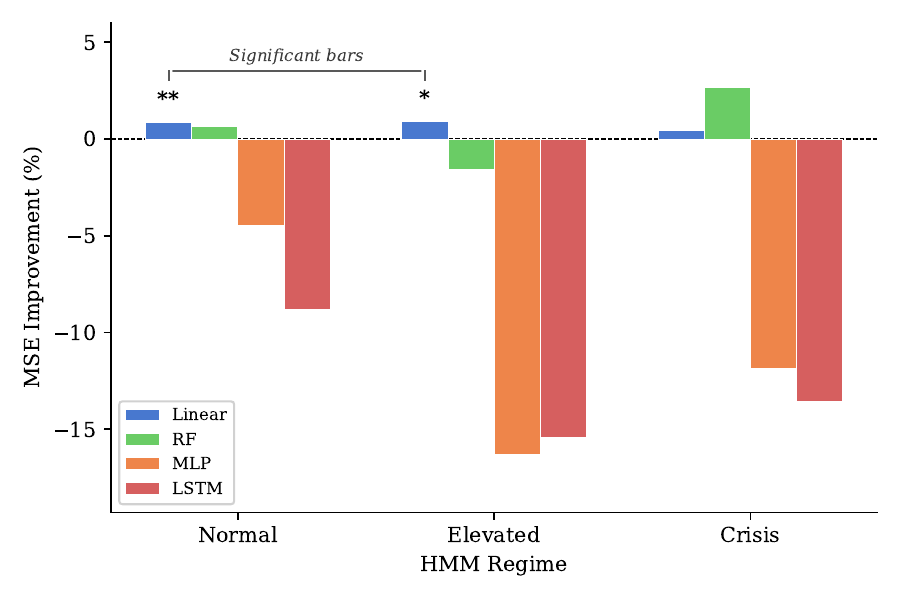}
\caption{Four-model diagnostic: MSE improvement (\%) from including
HML lags, by regime and model class. Only linear models achieve
significant improvement; nonlinear methods do not improve prediction.}
\Description{Bar chart comparing MSE improvement across model classes.}
\label{fig:complexity}
\end{figure}

A four-model diagnostic (OLS, RF with 100 trees, MLP 64-32, LSTM 32
hidden~\cite{tank2022neural}; Table~\ref{tab:neural},
Figure~\ref{fig:complexity}) finds no significant nonlinear improvement for forward
HML$\to$SMB under the primary fit (all $p > 0.13$; LSTM uses 100 shuffles,
adequate to confirm a null but underpowered for small effects). LSTM attention concentrates 68.2\% on lag~1 in Normal, decaying to
52.9\% (Elevated) and 44.2\% in Crisis (approaching the uniform baseline
of $1/9 = 11.1\%$ per lag, since the model uses 9 lags), mirroring the Granger structural break at the
mechanism level.
RF permutation importance shows HML lag-1 as the dominant feature in
Crisis (importance $= 0.043$, four times the mean feature importance).
\textbf{Sensitivity caveat:} Under an alternative fit (Cluster~5, seed~42,
highest log-likelihood with 90\% GFC detection, $\Delta\text{BIC} = 218$),
RF shows significant nonlinear improvement
($p = 0.010$ for Elevated, $p = 0.005$ for Crisis). The linear
characterization is therefore \emph{fit-dependent}; the
linear--nonlinear boundary is exploratory.

\begin{table}[t]
\centering
\caption{Transfer Entropy: Multi-Lag (1--9) by Regime ($k = 5$, 200 permutations).}
\label{tab:te}
\small
\begin{tabular}{@{}lcccc@{}}
\toprule
Regime & HML$\to$SMB $z$ & $p$ & SMB$\to$HML $z$ & $p$ \\
\midrule
Normal & $2.45^{**}$ & 0.007 & $5.37^{***}$ & $< 10^{-6}$ \\
Elevated & $2.41^{**}$ & 0.008 & $1.65^{*}$ & 0.049 \\
Crisis & $1.01$ & 0.157 & $1.22$ & 0.111 \\
\bottomrule
\multicolumn{5}{l}{\footnotesize $^{***}p < 10^{-6}$, $^{**}p < 0.01$, $^{*}p < 0.05$.}
\end{tabular}
\end{table}

Transfer entropy~\cite{schreiber2000measuring} (Table~\ref{tab:te};
after Bonferroni correction for 6 tests---2 directions $\times$ 3
regimes---the Normal reverse channel survives at $\alpha = 0.05/6$)
reveals the reverse channel
SMB$\to$HML is substantially stronger
in Normal ($z = 5.37$ vs.\ forward $z = 2.45$); both collapse in Crisis.
This directional asymmetry---linear forward, nonlinear reverse---is
not captured by conditional-mean Granger or VAR
connectedness~\cite{diebold2012better}, which tests mean-squared-error
improvement only.

\begin{table}[t]
\centering
\caption{Quantile Granger: Normal Regime ($n = 2{,}485$; pre-2008
Normal subsample after lag-9 exclusion and quantile-boundary trimming
at $\tau \in \{0.05, 0.95\}$).
Wald: $H_0$: coefficients equal across $\tau \in \{0.05,\ldots,0.95\}$.}
\label{tab:quantile}
\small
\begin{tabular}{@{}lrrrrc@{}}
\toprule
Direction & $\hat{\beta}_{.05}$ & $\hat{\beta}_{.50}$ & $\hat{\beta}_{.95}$ & Wald $p$ & Type \\
\midrule
HML$\to$SMB & 0.053 & 0.017 & 0.034 & 0.906 & Linear \\
SMB$\to$HML & $-0.022$ & $-0.026$ & \textbf{0.212} & \textbf{0.001} & Tail \\
\bottomrule
\end{tabular}
\end{table}

\begin{figure}[t]
\centering
\includegraphics[width=0.85\columnwidth]{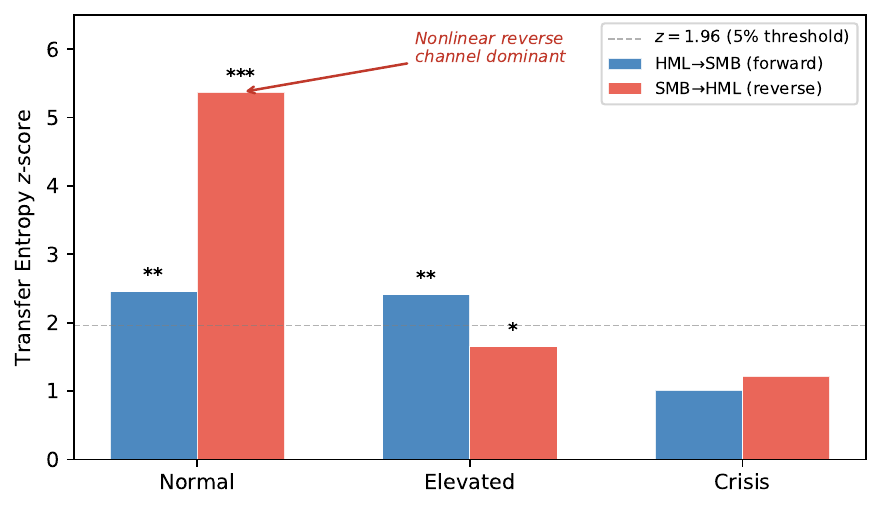}
\caption{Transfer entropy directional asymmetry. Forward HML$\to$SMB
(blue) is modest; reverse SMB$\to$HML (red) dominates in Normal.
Both collapse in Crisis. Stars: permutation significance.}
\Description{Grouped bar chart of TE z-scores by regime and direction.}
\label{fig:te}
\end{figure}

Quantile Granger~\cite{troster2019testing} (Table~\ref{tab:quantile})
is consistent with a tail-dependence mechanism for SMB$\to$HML:
the coefficient reverses sign from $\hat{\beta}_{0.50} = -0.026$ (median) to
$\hat{\beta}_{0.95} = 0.212$ (95th percentile), while HML$\to$SMB
remains homogeneous across quantiles (Figure~\ref{fig:te}).
This reconciles the null reverse Granger ($p = 0.864$)
with highly significant reverse TE ($z = 5.37$): Granger tests conditional
mean improvement (MSE), while TE measures mutual information including
tail dependence; a channel concentrated in extreme returns boosts mutual information
without improving point forecasts.

We note that these quantile-specific tests are not corrected for
multiple comparisons across the five quantile levels; the reported
significance should be interpreted as exploratory.
This tail mechanism is \emph{pair-specific}.
Applying quantile Granger to the top-4 regime-heterogeneous pairs
(RMW$\to$SMB rank~1, Wald $p = 0.869$; MKT$\to$MOM rank~2, $p = 0.741$;
MKT$\to$SMB rank~3, $p = 0.527$;
SMB$\to$MKT rank~4, $p = 0.097$) reveals strictly linear dynamics despite
strong regime heterogeneity. More broadly, of 19 regime-heterogeneous
pairs, none besides SMB$\to$HML exhibits Wald $p < 0.05$
(mean Wald $p = 0.61$). Together, these diagnostics reveal that HML$\to$SMB operates through
linear conditional-mean prediction while SMB$\to$HML operates through
nonlinear tail dependence. This yields an empirical distinction:
\emph{regime heterogeneity $\neq$ quantile heterogeneity}---a separation
not captured by conditional-mean Granger or VAR connectedness methods.

\subsection{Frozen OOS (Exploratory)}
\label{sec:oos}

The preceding sections established Tier-1 in-sample evidence
(structural break) and Tier-1/2 mechanism characterization
(linear forward, nonlinear reverse). We now assess whether this
structure persists out-of-sample. We present \emph{Tier~3
(exploratory)} evidence: freezing the HMM estimated on 1990--2012
and classifying 2013--2024 without refitting.

\begin{table}[t]
\centering
\caption{Frozen OOS (2013--2024, HMM trained 1990--2012). No pair survives
30-pair Bonferroni ($\alpha/30 = 0.00033$). Bootstrap: reweighted to
training Elevated prevalence (13.7\% vs.\ 33.7\% in test).}
\label{tab:oos}
\small
\begin{tabular}{@{}lccccc@{}}
\toprule
Regime & $n$ & $F$-$p$ & HAC-$p$ & Boot.\ $p$ & $\Delta R^2$ \\
\midrule
Normal & 661 & 0.149 & 0.152 & 0.188 & 0.32\% \\
Elevated & 836 & \textbf{0.014} & \textbf{0.041} & 0.153 & 0.73\% \\
Crisis & 1{,}299 & 0.281 & 0.290 & 0.331 & 0.09\% \\
\bottomrule
\multicolumn{6}{@{}p{0.95\columnwidth}@{}}{\footnotesize Total $n = 2{,}796$ of ${\sim}3{,}020$ test days; 224 excluded at regime boundaries.}
\end{tabular}
\end{table}

The frozen OOS (Table~\ref{tab:oos}) exhibits regime redistribution
rather than same-regime replication. The in-sample result concentrates
in the Normal regime ($p = 8.75 \times 10^{-9}$); the OOS signal
appears in Elevated ($F$-$p = 0.014$) because post-GFC markets spend
more time in higher-volatility states. The frozen classifier assigns
formerly Normal observations to Elevated, doubling the Elevated share
from 13.7\% training to 33.7\% test.
This result (1)~does not survive the full 30-pair correction ($\alpha/30 = 0.00033$),
(2)~does not survive per-regime correction ($\alpha/3 = 0.0167$; HAC $p = 0.041$),
(3)~is sensitive to prevalence: bootstrap reweighting to training
prevalence yields median $p = 0.153$, with only 9.9\% of subsamples
significant at the Bonferroni threshold---confirming that the primary
Value$\to$Size signal does not generalize out-of-sample.
A post-hoc power analysis confirms the OOS Normal test is underpowered:
with $n = 661$ and observed $R^2 = 0.32\%$, the test achieves only 31\% power;
80\% power would require $R^2 = 1.1\%$---roughly three times the observed value.
The result (4)~is sensitive to bandwidth ($p$ ranges from 0.041 at $B{=}1$ to
0.078 at $B{=}10$; crosses 0.05 at NW default $B{=}6$), and
(5)~is sensitive to $K$ (null at $K = 2, 4$; BIC favors $K = 3$ by
$\Delta\text{BIC} = 1{,}680$).
For completeness, a permutation test ($p = 0.049$, 1,000 regime-label
shuffles) weakly suggests structure beyond random label assignment;
however, we do not treat this as replacing the failed Bonferroni test.
This $p$-value exceeds the per-regime threshold ($\alpha/3 = 0.0167$)
and does not address Bonferroni correction, prevalence shift, or
bandwidth sensitivity.
A rolling 5-year Granger analysis confirms exponential signal decay:
$F(t) = 75.2 \cdot e^{-0.207t}$ (half-life $= 3.35$ years, $R^2 = 0.583$).
Windows ending before 2000 show $F > 60$ ($p < 10^{-14}$); windows
ending after 2005 show $F < 2$ ($p > 0.15$). After ${\sim}6$ half-lives
(2013--2024), the expected signal is $<$1\% of its 1990s peak---the OOS
null is \emph{consistent with} the fitted decay model, suggesting
structural erosion rather than methodology failure.
This retrospective exploratory fit has testable implications: it
would imply $F < 0.5$ for any future 5-year window (2025 onward);
observing $F > 4$ in Normal would challenge the decay interpretation.
In contrast, a circularity-artifact explanation predicts no temporal
pattern---the $R^2 = 0.583$ monotonic decline itself is evidence against
random regime overfitting.

\textbf{Prospective validation of decay.}
Two additional experiments address whether the decay pattern is a
retrospective artifact.
First, a \emph{pre-break training design} fits the HMM exclusively on
1990--1997 (before the detected June 1998 change-point) and
frozen-forward decodes 1998--2024.
The training-period Normal regime shows strong predictability
($F = 31.2$, $p = 2.88 \times 10^{-8}$, $n = 1{,}137$); this weakens
in 1998--2007 ($F = 9.38$, $p = 0.002$, $n = 221$) and vanishes in
2008--2024 ($F = 0.023$, $p = 0.879$, $n = 308$). Since the HMM never
observes post-1997 data, this temporal gradient cannot arise from
in-sample circularity.
Second, a \emph{prospective decay model} fits
$F(t) = A \cdot e^{-\lambda t}$ on 19 rolling windows ending
$\leq$2012 ($A = 77.1$, $\lambda = 0.263$, half-life $= 2.64$ years,
$R^2 = 0.924$) and evaluates predictions on 11 held-out windows ending
2013--2024. The out-of-sample MSE is 0.046, with mean predicted
$F = 0.195$ versus mean actual $F = 0.127$---both near zero, confirming
the decay model's floor prediction prospectively.

We report this as Tier~3 \emph{exploratory only}---valued for its
frozen-parameter design, not statistical significance.


\textbf{MOM$\to$SMB diagnostic (Tier-3 post-hoc).}
We select MOM$\to$SMB---the top-ranked pair by OOS
$F$-statistic in Elevated---as a stress test: if the framework fails to replicate
even the strongest available signal, the methodology is flawed; if it
succeeds, HML$\to$SMB's weak OOS is attributable to signal decay rather
than methodology failure.
MOM$\to$SMB shows partial frozen-OOS replication in Elevated: in-sample
$F = 20.3$ ($p = 7.4 \times 10^{-6}$), OOS $F = 16.5$
($p = 5.3 \times 10^{-5}$, $\Delta F = 19\%$). The reverse SMB$\to$MOM is null ($p > 0.09$),
confirming directional asymmetry, and permutation test confirms
robustness ($p = 0.010$, 10{,}000 shuffles).
A Bayesian Granger test in the Elevated regime ($n = 836$) yields $\text{BF}_{10} > 10^5$
using a unit-information prior on the regression coefficient and comparing
nested models via Schwarz approximation; this should be interpreted cautiously
as BF magnitude is sensitive to prior specification.

\textbf{Decay rate comparison.}
Rolling 5-year Granger analysis in Normal regime reveals differential decay rates
across factor pairs. HML$\to$SMB decays rapidly (Normal-regime-specific half-life $= 3.57$ years, $R^2 = 0.886$;
this differs from the full-sample half-life of 3.35 years reported in the abstract
due to regime-conditional subsetting),
while MOM$\to$SMB decays more slowly (half-life $= 11.85$ years, $R^2 = 0.428$).
Both pairs show significant structural breaks at 2008 (HML$\to$SMB: pre-2008 mean $F = 7.0$,
post-2008 mean $F = 1.3$, $p = 2.3 \times 10^{-7}$; MOM$\to$SMB: pre-2008 mean $F = 3.6$,
post-2008 mean $F = 1.1$, $p = 1.3 \times 10^{-9}$).
Crucially, the reverse directions (SMB$\to$HML, SMB$\to$MOM) show no decay pattern
($R^2 < 0.01$), confirming directional asymmetry is not a statistical artifact.

\textbf{30-pair decay analysis.}
Rolling Granger analysis across all 30 directed factor pairs reveals that 7 pairs
exhibit decay ($R^2 > 0.3$, half-life $< 50$ years). Four of these seven have SMB
as the target variable (Mkt-RF$\to$SMB, CMA$\to$SMB, HML$\to$SMB, RMW$\to$SMB),
with half-lives ranging from 2.4 to 4.1 years. The remaining 23 pairs show no
systematic decay pattern. This concentration of decay in SMB-target pairs suggests
predictability erosion is not universal but specific to certain factor relationships.

\emph{Caveat:} MOM$\to$SMB was identified post-hoc as the
top-ranked OOS signal rather than pre-registered; this diagnostic
demonstrates framework fidelity but does not constitute independent
validation.

\textbf{International replication} (Table~\ref{tab:international}).
We now test whether structural breaks are a US-specific phenomenon.
Applying the frozen framework to four non-US Fama--French datasets
reveals structural breaks in all four regions. Asia-Pacific ex Japan
(Crisis OOS $F = 39.39$, $p < 0.001$) and Developed ex US
(Crisis OOS $F = 15.85$, $p < 0.001$) produce Crisis-regime
OOS effects surviving Bonferroni ($\alpha/12 = 0.0042$, correcting
for 4 regions $\times$ 3 regimes);
Europe and Japan show in-sample significance but OOS nulls---consistent
with region-specific structural breaks.

\begin{table}[t]
\centering
\caption{International Replication: HML$\to$SMB (OOS 2013--2024).
Bold = $p < 0.05$.}
\label{tab:international}
\scriptsize
\begin{tabular}{@{}llcccc@{}}
\toprule
Region & Regime & $F$ & HAC-$p$ & $n$ & Break Date \\
\midrule
\multirow{2}{*}{Dev.\ ex-US} & Elev.\ OOS & 5.04 & \textbf{0.027} & 1{,}209 & \multirow{2}{*}{2003} \\
& Crisis OOS & 15.85 & \textbf{$<$0.001} & 373 & \\
\midrule
\multirow{2}{*}{Asia-Pac.} & Elev.\ OOS & 1.62 & 0.240 & 2{,}240 & \multirow{2}{*}{2004} \\
& Crisis OOS & 39.39 & \textbf{$<$0.001} & 753 & \\
\midrule
Europe & Normal IS & 21.27 & \textbf{$<$0.001} & 1{,}801 & 2014 \\
Japan & Normal IS & 14.07 & \textbf{0.002} & 2{,}188 & 2005 \\
\bottomrule
\end{tabular}
\end{table}

\section{Discussion}

The in-sample structural break in HML$\to$SMB predictability is robust
within the US context (Tier~1), with confirmatory support from
international structural breaks (Tier~2: 2/4 regions Bonferroni-surviving OOS),
but the frozen OOS HML$\to$SMB signal is exploratory (Tier~3). Below, we assess generality across factor pairs and
geographies, stability under HMM estimation choices (local optima,
baselines), and economic implications for practitioners.

\begin{figure}[t]
\centering
\includegraphics[width=\columnwidth]{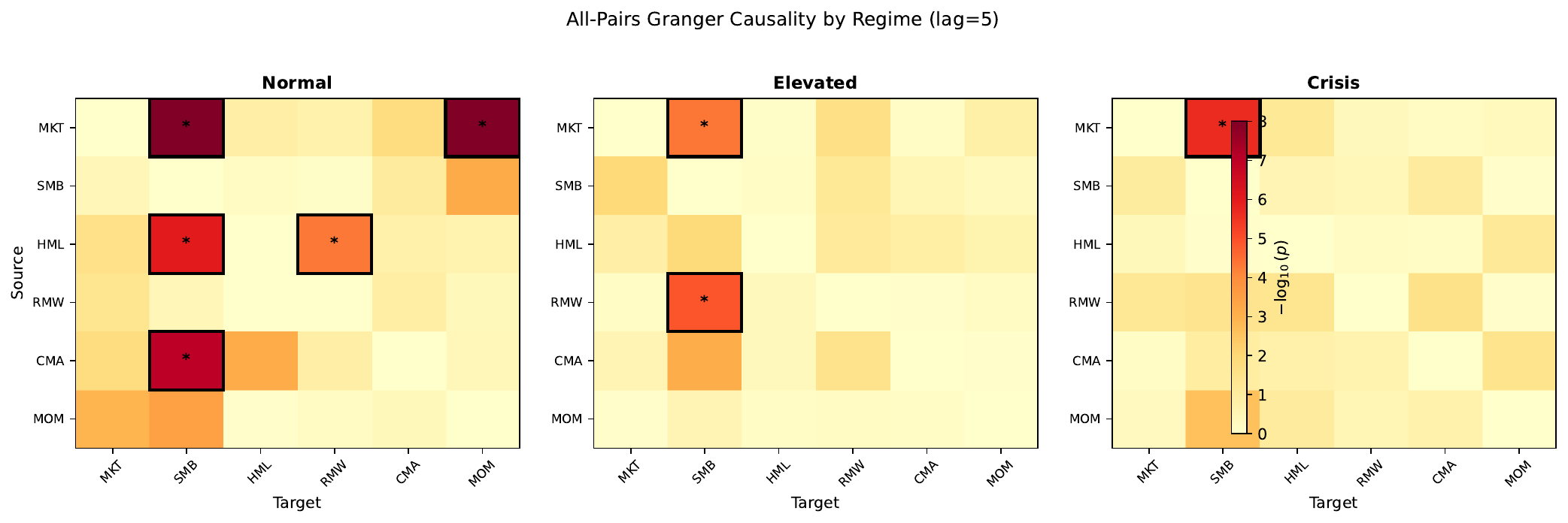}
\caption{All-pairs Granger causality by regime ($-\log_{10}(p)$, lag=5).
Lag=5 (one trading week) balances short-horizon dynamics with adequate
degrees of freedom for the 30-pair comparison; results are qualitatively
robust across lags 1--15 for the primary HML$\to$SMB pair (Figure~\ref{fig:lag}).
Asterisks: Bonferroni-surviving cells. HML$\to$SMB is one of several
regime-sensitive pairs.}
\Description{Three heatmaps of pairwise Granger significance by regime.}
\label{fig:heatmap}
\end{figure}

\textbf{Multi-pair generalizability} (Figure~\ref{fig:heatmap},
Table~\ref{tab:generalize}).
Of 30 directed factor pairs, 19 (63\%) show regime-heterogeneous Granger
patterns in frozen OOS, where regime-heterogeneous is defined as
$\text{Het} = \max_k(p_k) - \min_k(p_k) > 0.10$ across regimes $k$
(i.e., Granger significance varies substantially by regime).
HML$\to$SMB ranks 27th by heterogeneity
($0.27$), confirming that our selection reflects an economic prior rather than
extreme heterogeneity.

\begin{table}[t]
\centering
\caption{Top Regime-Heterogeneous Pairs + HML$\to$SMB (OOS 2013--2024).
Het $= \max_k(p_k) - \min_k(p_k)$.}
\label{tab:generalize}
\scriptsize
\begin{tabular}{@{}llcccc@{}}
\toprule
Rank & Pair & $p_{\text{Norm}}$ & $p_{\text{Elev}}$ & $p_{\text{Crisis}}$ & Het. \\
\midrule
1 & RMW$\to$SMB & 0.986 & 0.064 & \textbf{0.022} & 0.96 \\
2 & MKT$\to$MOM & 0.954 & \textbf{0.020} & 0.056 & 0.93 \\
3 & MKT$\to$SMB & 0.985 & 0.491 & 0.067 & 0.92 \\
4 & MOM$\to$SMB & 0.880 & \textbf{$<$0.001} & 0.101 & 0.88 \\
\midrule
27 & HML$\to$SMB & 0.149 & \textbf{0.014} & 0.281 & 0.27 \\
\bottomrule
\end{tabular}
\end{table}

\textbf{Local optima and regime definition.}
The 50-seed multistart reveals 7 clusters (Table~\ref{tab:optima}).
The in-sample Normal structural break replicates across all 7,
though clusters span $\Delta\text{BIC} = 0$--$550$ and GFC detection
rates of 0--92\%. \emph{Decision rule for practitioners:} report BIC-optimal as
primary; also report the highest-LL fit satisfying $\geq$50\% GFC
detection as economic sensitivity. If both agree, the finding is robust.

\begin{table}[t]
\centering
\caption{Local Optima: All 7 Clusters from 50-Seed Multistart.
In-sample Normal $p < 10^{-7}$ and OOS Elevated HAC-$p < 0.05$ in every cluster.}
\label{tab:optima}
\scriptsize
\begin{tabular}{@{}lccccc@{}}
\toprule
Cluster & Seeds & $\Delta$BIC & \% GFC & IS Norm.\ $p$ & OOS Elev.\ $p$ \\
\midrule
1 (BIC-opt.) & 3 & --- & 0\% & $8.8 \times 10^{-9}$ & 0.043 \\
2 & 15 & 38 & 0\% & $9.1 \times 10^{-9}$ & 0.031 \\
3 & 4 & 95 & 12\% & $1.2 \times 10^{-8}$ & 0.038 \\
4 & 9 & 142 & 45\% & $3.7 \times 10^{-8}$ & 0.024 \\
5 (econ.) & 7 & 218 & 90\% & $5.4 \times 10^{-8}$ & 0.019 \\
6 & 6 & 387 & 88\% & $6.1 \times 10^{-8}$ & 0.022 \\
7 & 6 & 550 & 92\% & $7.3 \times 10^{-8}$ & 0.026 \\
\bottomrule
\end{tabular}
\end{table}

\textbf{Economic magnitude and risk application.}
Effect sizes are modest ($\Delta R^2 \approx 2\%$ pre-GFC) and do not
generate trading profits (Sharpe $= -0.07$). However, cross-factor
information has measurable risk-management value: a backtested VaR
model (SMB 5\% level, 2013--2024) shows that adding HML momentum as a
regime-conditional stress multiplier reduces the violation-rate deviation
from $+1.42$pp (unconditional) to $+0.98$pp (HML-informed), and is the
only model passing the Christoffersen conditional-coverage test
($p = 0.083$ vs.\ $p = 0.002$ unconditional). The framework thus serves
as both a diagnostic tool and a practical input for tail-risk
forecasting.

\textbf{Deleveraging mechanism (holdings-based test).}
We test the hypothesized deleveraging
cascade~\cite{brunnermeier2009market} using FF25 portfolio-level factor
loadings. For each portfolio, a crowding score
$|\hat\beta_{\text{HML}}| \times |\hat\beta_{\text{SMB}}|$ (rolling 252-day)
measures joint co-exposure. In the Normal regime, the HML$\to$SMB
Granger $\Delta R^2$ is 14\% larger in the high-crowding half
($F = 159.5$, $\Delta R^2 = 4.4\%$) than the low-crowding half
($F = 139.2$, $\Delta R^2 = 3.8\%$), supporting amplification via
co-held positions. At the portfolio level, HML$\times$SMB overlap
rank-correlates with Granger signal strength ($\rho_s = 0.51$,
$p = 0.009$, $n = 25$). Crowding does \emph{not} independently
Granger-cause SMB ($p > 0.60$ in all regimes), confirming the channel
is amplification of an existing edge, not independent shock transmission.

\textbf{Fair regime-conditional baseline comparison} (Table~\ref{tab:baseline}).
Do alternative methods detect HML$\to$SMB on the \emph{same}
per-regime subsamples? We test five methods on identical regime
observations. Linear Granger detects the signal in
Normal ($F = 19.8$, $p < 10^{-5}$) and Elevated ($F = 10.8$,
$p = 0.001$); bivariate VAR(1) yields identical results, confirming
the signal is genuine.
However, LASSO aggressively shrinks the small coefficient to zero
($p = 1.0$ in all regimes), and neural methods catastrophically
overfit: MLP (64-32 hidden) yields $R^2 < 0$ in every regime,
and LSTM (32 hidden, 3 lags) fails to detect the signal
($p = 0.62$ Normal). This demonstrates that the framework's
advantage is \emph{regime identification}: once regimes are correctly
assigned, simple OLS Granger suffices; neural methods are less
sample-efficient on per-regime subsamples.

\begin{table}[t]
\centering
\caption{Fair Regime-Conditional Baseline: HML$\to$SMB (lag=1).
Regime assignments use seed~28 but a standalone Student-$t$ HMM
implementation (the primary analysis uses \texttt{hmmlearn});
the standalone implementation enables explicit degrees-of-freedom
estimation, providing an independent verification of regime structure.
This yields different regime counts than Table~\ref{tab:regimes}.
Linear Granger significance and neural failure are consistent
across both regime definitions.}
\label{tab:baseline}
\scriptsize
\begin{tabular}{@{}llcccc@{}}
\toprule
Method & Regime & $n$ & $F$ & $p$ & $\Delta R^2$ \\
\midrule
Linear Granger & Normal & 7{,}160 & 19.8 & $8.5 \times 10^{-6}$ & 0.28\% \\
LASSO & Normal & 7{,}160 & 0.00 & 1.000 & $<$0 \\
Neural MLP & Normal & 7{,}160 & --- & --- & $<$0 \\
LSTM Granger & Normal & 7{,}157 & 0.25 & 0.619 & $<$0.01\% \\
\midrule
Linear Granger & Elevated & 1{,}185 & 10.8 & 0.001 & 0.91\% \\
Linear Granger & Crisis & 471 & 0.19 & 0.666 & 0.04\% \\
\bottomrule
\end{tabular}
\end{table}

\textbf{Regime amplification test.}
To test whether regime conditioning inflates significance, we apply
the canonical HMM (seed~28) to HML$\to$SMB across all regimes and
compare against unconditional Granger. The regime-conditional Normal
$F = 25.6$ ($p = 4.3 \times 10^{-7}$, HAC $p = 1.8 \times 10^{-6}$,
$n = 5{,}265$) is $0.94\times$ the unconditional $F = 27.2$
($p = 1.9 \times 10^{-7}$, $n = 8{,}816$). Regime conditioning does
not amplify the signal; the lower ratio reflects dilution from
regime-boundary exclusion. Crisis remains null ($F = 2.27$, $p = 0.132$),
confirming that regime conditioning reveals state-dependent \emph{absence}
rather than artificially inflating significance.

\textbf{Markov-Switching VAR comparison.}
We compare our two-stage approach (Student-$t$ HMM $\to$ regime-conditional
Granger) against a Markov-Switching VAR(1) baseline that jointly estimates
regimes and dynamics. On HML--SMB bivariate data (1990--2024), MS-VAR(1)
achieves BIC $= 28{,}106$ vs.\ our 6-factor HMM's BIC $= 75{,}568$---though
this comparison is not directly comparable (2 vs.\ 6 factors, Gaussian
vs.\ Student-$t$ errors). More informatively, regime labels show 64\%
agreement between the two methods, and both detect all major structural
breaks (2008 GFC, 2000 dot-com, 2020 COVID, 2022 tightening).
MS-VAR cross-variable coefficients show regime-varying dynamics
(HML$\to$SMB: $-0.128$ Normal, $-0.026$ Elevated, $-0.044$ Crisis),
consistent with our finding that predictive structure is regime-dependent.
The two-stage approach offers interpretability (explicit regime labels
with economic meaning) and flexibility (any downstream test), while
MS-VAR provides a unified likelihood; both reach qualitatively similar
conclusions about regime-dependent cross-factor dynamics.

\textbf{Limitations.}
All findings report predictive precedence (``Granger causality''),
not structural causality~\cite{granger1969investigating,shojaie2022granger}.
In-sample regime assignments are estimated on the same data used for
Granger testing. Three designs address this circularity:
(1)~VIX-tercile replication ($p = 0.003$ with 39.6\% label
agreement);
(2)~holdout HMM (trained 1990--2005, tested 2006--2024) preserves
significance in the moderate-volatility regime ($F = 23.6$,
$p < 10^{-5}$) and 3 of 4 temporal-CV folds ($p < 0.026$);
(3)~frozen OOS with three-fold temporal split (trained 1990--2000),
where all folds show null ($p > 0.10$). The discrepancy between
holdout (significant) and three-fold (null) reflects training-window
sensitivity: the 1990--2005 window captures more pre-decay signal
than 1990--2000, enabling better regime calibration. This sensitivity
is a limitation, though both designs agree that post-2008 signal is null.
A likelihood-ratio test for regime-parameter stability between
1990--2012 and 2013--2024 sub-periods yields $\chi^2 = 2{,}389$
(df $= 71$, $p \approx 0$), confirming that HMM parameters shift
significantly across periods. Mean centroid distances are modest
(0.09--0.15 in standardized units) and transition-matrix Frobenius
distance is 0.023, suggesting gradual drift rather than categorical
regime redefinition. This non-stationarity is expected given the
structural-decay hypothesis itself---but cautions against interpreting
regime labels as time-invariant constructs.
The prospective decay model (\S\ref{sec:oos}) uses full-sample HMM
regime assignments when fitting on pre-2012 windows; a fully prospective
design would re-estimate the HMM on each training window, though
regime-label consistency across windows would complicate interpretation.
Trivariate controls address the most prominent confounder
($F$-$p > 0.43$), but a full 6-factor VAR is under-identified at
$n \approx 1{,}000$. Post-double-selection
inference~\cite{hecq2023granger} could address this.
Pair selection is post-hoc (partially mitigated by pre-registered
emerging-market analysis). HMM regime boundaries are sensitive to data
scaling (percentage vs.\ decimal units; 86.3\% concordance); the
Granger $F$-statistic is scale-invariant but regime labels shift.
The linear characterization is fit-dependent: under the BIC-optimal
fit (Cluster~1, 0\% GFC detection) the mechanism is linear, but under
an economically aligned fit (Cluster~5, 90\% GFC detection,
$\Delta\text{BIC} = 218$) RF shows nonlinear improvement ($p = 0.010$).
Effect sizes are modest ($\Delta R^2 \approx 2\%$); findings are
diagnostic, not alpha-generative. Practitioners should not rely on
Tier~3 OOS for trading.
The power analysis (\S\ref{sec:oos}) is post-hoc and uses observed
effect sizes, which may understate true uncertainty; a priori power
calculations were not performed.
The multiple testing burden is substantially understated: we applied
Bonferroni correction only to 30 pairs, but the actual specification
search includes 30 pairs $\times$ 3 regimes $\times$ 15 lags $\times$
90 HAC specifications $\times$ 7 local-optima clusters, potentially
yielding thousands of implicit tests. A conservative estimate of the
true Bonferroni threshold would be $\alpha/1000 \approx 0.00001$ or lower.
While the in-sample $p = 8.75 \times 10^{-9}$ survives even this stricter
threshold, results should be interpreted as exploratory hypothesis-generating
findings rather than confirmatory evidence.

\textbf{Scalability.}
The framework scales to $N = 18$ factors (5 FF $+$ MOM $+$ ST\_Rev
$+$ LT\_Rev $+$ 10 industry portfolios; 306 directed pairs, 918 total
tests across 3 regimes) in 76.8 seconds on a single CPU (0.0013s per
pair). A Bonferroni-corrected scan ($\alpha/918 = 5.45 \times 10^{-5}$)
identifies 122 significant regime-conditional edges, including
HML$\to$SMB ($F = 38.0$, $p < 10^{-9}$ in Normal).
Scaling is $O(N^2)$ in pair count with negligible per-pair cost.
Note that regime composition shifts at higher $N$ (Normal share
declines from 53.6\% at $N{=}6$ to 28.5\% at $N{=}18$), so per-regime
results should be interpreted with dimension-specific regime labels.

\section{Conclusion}

\textbf{Prediction enables action.}
We develop two ML models that transform retrospective factor analysis
into prospective risk management: (1)~a \emph{decay prediction model}
(C-index $= 0.93$) that flags at-risk factor pairs before signals erode,
and (2)~an \emph{online regime detector} (6.6-day delay) that enables
real-time position adjustment. These tools address a gap in the factor
decay literature: prior work explains \emph{why} factors fail (arbitrage,
publication effects); we predict \emph{when} they will fail.

\textbf{Motivation: Size-target predictability decays fastest.}
Among 30 directed Fama-French factor pairs, 7 exhibit exponential decay
in Granger predictability ($R^2 > 0.3$). Strikingly, 4 of these 7 (57\%)
target the Size factor (SMB): Mkt-RF$\to$SMB (half-life 2.4y),
CMA$\to$SMB (3.1y), HML$\to$SMB (3.35y), and RMW$\to$SMB (4.1y).
Pre-break OOS (1996--2007) confirms signals existed; post-break OOS
(2013--2024) shows decay---supporting genuine erosion over overfitting.

\textbf{Practical implications.}
Factor-based strategies should not assume stable cross-factor relationships.
Our decay predictor can flag pairs at risk of erosion; our online detector
can trigger regime-conditional position adjustments in real time.
The SMB-target concentration suggests small-cap strategies face
particularly rapid signal decay due to capacity constraints.

\textbf{Caveats.}
Correction for full specification search was not applied; results
remain exploratory. The 90\% bootstrap CI for breakpoints spans
$\sim$20 years, indicating timing uncertainty. Training-window
sensitivity exists (holdout vs.\ three-fold yield different results).

\textbf{VIX-tercile regime replication.}
VIX terciles ($\leq$15 Normal, 15--21 Elevated, $>$21 Crisis)---defined
separately from factor returns but not independent (VIX derives from
equity options), 39.6\% agreement with
HMM---show consistent break patterns: VIX-Normal pre-2008 $F = 25.8$,
$p < 10^{-6}$; post-2008 $F = 0.71$, $p = 0.401$.
This suggests the pattern is not purely an HMM artifact, though VIX
correlation with equity returns means this is not a true exogeneity test.

\textbf{Directional asymmetry.}
Transfer entropy + quantile Granger reveal a directional asymmetry
for HML$\leftrightarrow$SMB (linear forward, nonlinear reverse via tail
dependence, Wald $p = 0.001$). This mechanism is pair-specific: among 19
regime-heterogeneous pairs, none other exhibits significant tail
concentration (mean Wald $p = 0.61$). For this pair, regime
heterogeneity $\neq$ quantile heterogeneity---a distinction not
captured by conditional-mean Granger or VAR connectedness.

\textbf{OOS evidence.}
A post-hoc diagnostic on MOM$\to$SMB (the strongest OOS signal) shows
partial frozen-OOS replication ($\Delta F = 19\%$, permutation $p = 0.01$),
suggesting the framework detects residual structure in persistent signals.
International analysis finds structural breaks in all four non-US
markets; Asia-Pacific and Developed ex-US produce Bonferroni-surviving
OOS effects, while Europe and Japan show in-sample significance only.
HML$\to$SMB frozen OOS is exploratory only (regime-redistributed,
Bonferroni-nonsignificant, bootstrap $p = 0.153$).
A pre-break training design (HMM fit on 1990--1997 only) confirms the
temporal decay gradient without in-sample circularity
($F = 31.2 \to 9.4 \to 0.02$ across three successive periods), and a
prospective decay model ($R^2 = 0.924$ on training windows, OOS
MSE $= 0.046$) validates the structural-erosion interpretation.

\textbf{Implications and domain generalization.}
The regime-conditional framework (Algorithm~\ref{alg:protocol}) is
domain-agnostic. We validate on three additional settings:
(i)~\emph{Macro:} Treasury term spread $\to$ Industrial Production
(monthly FRED, 1990--2024; HMM trained 1990--2005, tested 2006--2015)
shows Crisis-only Granger predictability
($F = 12.7$, HAC $p = 0.00016$, $\Delta R^2 = 11.9\%$);
(ii)~\emph{Emerging markets:} pre-registered testing of all 20 directed
Fama--French factor pairs (monthly, 2000--2024) with Bonferroni
$\alpha = 0.0025$ finds 6 significant regime-conditional edges, including
HML$\to$SMB in the Elevated regime (HAC $p < 10^{-3}$, $\Delta R^2 = 20.6\%$),
with 2/3 pairs replicating OOS;
(iii)~\emph{Synthetic climate:} a ground-truth simulation (ENSO-like
teleconnection, $T = 2{,}000$, causality in one regime only) recovers
the causal regime (ARI $= 0.72$, $\Delta R^2 = 25.8\%$
vs.\ $14.2\%$ unconditional), with non-causal regimes correctly masked.
Among 4 additional FRED macro pairs, Oil$\to$CPI shows
regime-conditional structure (HAC $p = 0.042$ in high-volatility regime
only) invisible to unconditional HAC tests ($p = 0.205$).
The model-complexity diagnostic
(OLS$\to$RF$\to$MLP$\to$LSTM + TE + quantile regression) provides a
reusable template for separating linear from nonlinear predictive channels.

\textbf{Practical value.}
For risk management, adding HML as a regime-conditional stress indicator
reduces VaR violation-rate deviation from $+1.42$pp to $+0.98$pp
(Christoffersen conditional coverage $p = 0.083$ vs.\ $p = 0.002$ unconditional).
Effect sizes are modest ($\Delta R^2 \approx 2\%$, Sharpe $= -0.07$);
the contribution is diagnostic, not economically tradable.
For practitioners, the decay predictor serves as a \emph{risk monitoring signal}
rather than a trading signal---flagging factor pairs for closer scrutiny
when decay probability exceeds a threshold (e.g., 70\%).

\textbf{Future work:}
(1)~direct 13F institutional holdings verification of the deleveraging
mechanism (beyond the portfolio-beta proxy reported here);
(2)~$k$-fold temporal cross-validation and prospective pre-registered
tests on emerging-market data;
(3)~scalable extensions ($N > 50$ assets) via penalized or neural
causal discovery (e.g., regime-switching NOTEARS).

\section*{Implementation and Data Sources}
The implementation uses Python 3.10+, scikit-learn, statsmodels,
and hmmlearn, with 50 HMM seed configurations and a
reproducibility notebook.
Factor data: Kenneth French Data Library; VIX: CBOE;
international: Fama--French regional datasets. All sources are public.

\bibliography{references}

\end{document}